\documentclass[aps,prx,10pt,amsmath,amssymb,floatfix,notitlepage,twocolumn]{revtex4-2}

\usepackage{graphicx}   
\usepackage{dcolumn}    
\usepackage{bm}         

\begin{document}

\title{Resonant current from singlet-triplet state mixing in coupled quantum dots}

\author{G. Giavaras}\email{Quantum Laboratory Kivotos, g.giavaras@gmail.com}


\begin{abstract}
Electrically driven spin resonances in double quantum dots can
lift the spin blockade and give rise to a resonant current. This
current can probe the properties of coupled two-spin states for
different quantum dot configurations. Using a Floquet-Markov
quantum transport model we compute the resonant current for
different driving amplitudes and ac field frequencies in
spin-orbit coupled quantum dots. We show that the resonant current
has a very rich interference pattern which can give valuable
insight into the singlet-triplet state mixing.
\end{abstract}

\maketitle

\textit{Introduction--} Spins in semiconductor quantum dots can be
controlled electrically and are promising not only for
quantum-gate engineering~\cite{zalba} but also for studying
correlation effects, decoherence, and transport properties in the
nanoscale~\cite{elze, petta, ono17, georgios, abdu, xu, sala,
giavaras07a, nowak, xue, zhou, giavaras07b, giavaras16, li, wang,
giavarasE, cota, song, chorley11, zhang21, liu, giavaras13,
zhang23, giavaras10, minmin, ono19, giavaras19a, giavaras19b,
giavaras20}. In a double quantum dot (DQD) in the spin-blockade
regime~\cite{ono02} the singlet energy levels define a two-level
system with an anticrossing point that is the result of coherent
interdot tunneling. When an ac electric field is added to the
DQD~\cite{fujisawa2} this anticrossing allows for
Landau-Zener-St\"uckleberg-Majorana (LZSM) transitions and
interference~\cite{kohler, ryzhov, shevchenko1, shevchenko2}. Such
transitions have been studied in detail not only in
semiconducting~\cite{kohler, ryzhov, shevchenko1, shevchenko2} but
also in various superconducting quantum systems~\cite{shevchenko1,
shevchenko2}. Furthermore, hybridized singlet-triplet states in
tunnel-coupled quantum dots can serve as qubits which can be
probed with microwave fields and electrical transport
measurements~\cite{zalba, elze, petta, ono17, perge12,
giavaras19a, takahashi, kanai}. Then, some of the coherence
properties of the qubit can be extracted from the resonant current
flowing through the DQD.

For the appropriate driving amplitude the current can map out the
two-spin energy levels~\cite{ono17, perge12, giavaras19a,
giavaras19b, giavaras20} and the singlet-triplet anticrossing.
Depending on the energy configuration of the DQD the ac field can
induce different types of transitions. Some typical cases include,
for example, transitions between the two energy levels forming the
anticrossing or transitions between a third energy level and the
two levels that anticross. In both cases the ac field induced
resonant current has been measured revealing the singlet-triplet
anticrossing~\cite{perge12, ono17}. Here, we show that beyond the
spectroscopic regime which focuses on the anticrossing gap, the
resonant current can exhibit a rich interference pattern. This can
be very sensitive to the exact driving frequency and the resonant
region can undergo a drastic change. Exploring the current within
a narrow energy detuning range and different ac field frequencies
gives insight into a singlet-triplet state mixing.

\begin{figure}
\includegraphics[width=4.2cm, angle=270]{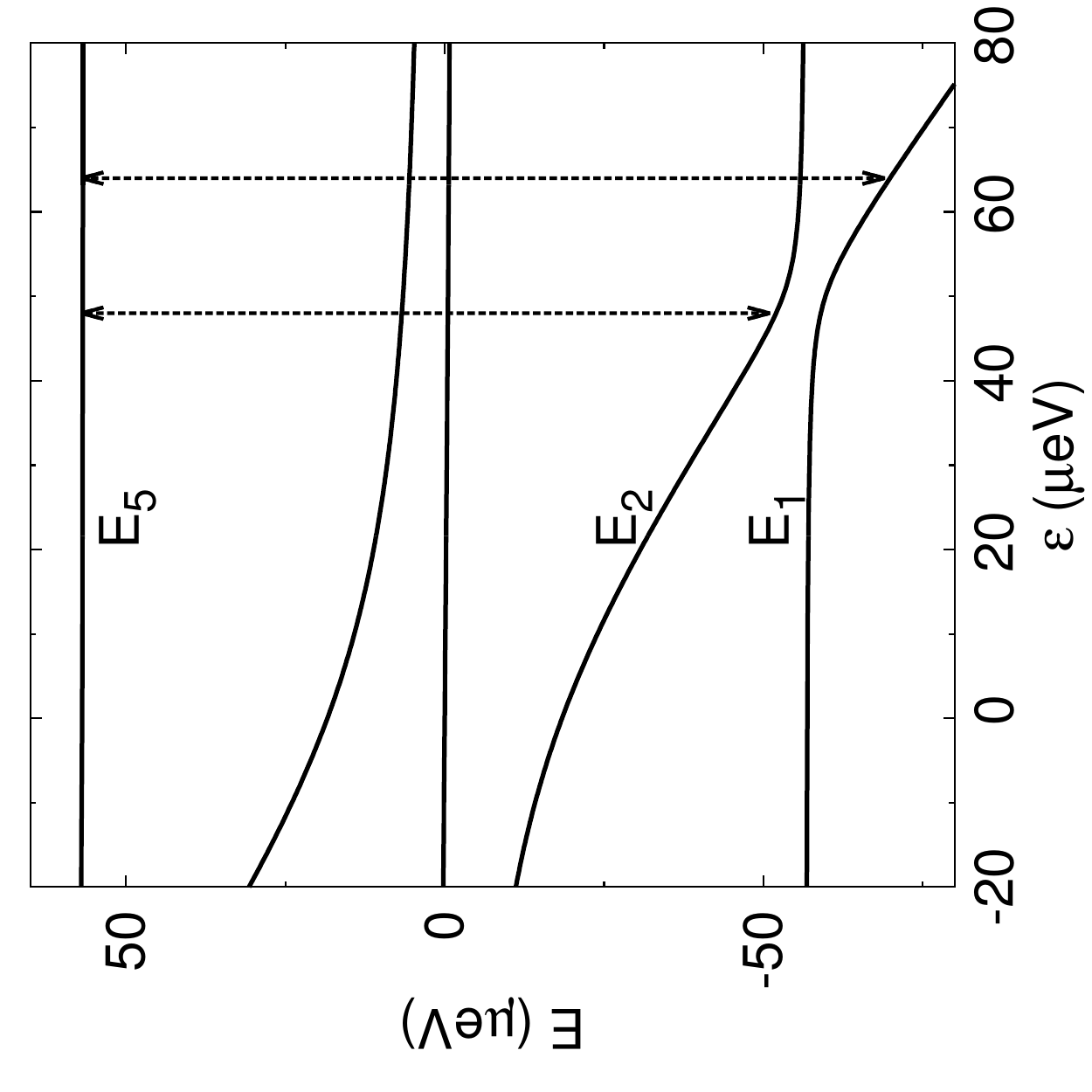}
\includegraphics[width=4.2cm, angle=270]{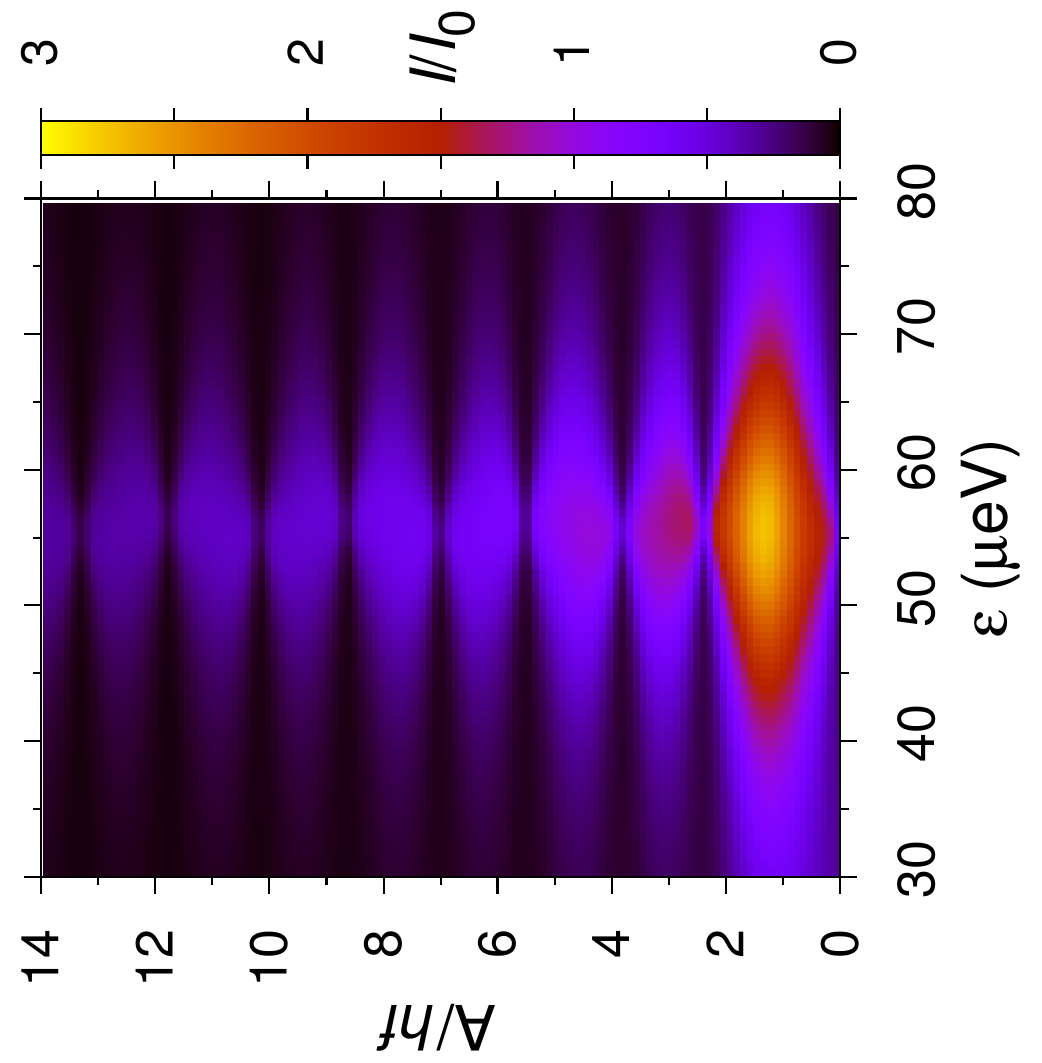}
\caption{Left frame: Energy levels of double quantum dot as a
function of energy detuning at $B=0.135$ T with $t_c=13$ $\mu$eV,
$t_{\rm so}=3.5$ $\mu$eV, $g_1=7$, and $g_2=7.5$. Typical
frequencies considered in this work are indicated by vertical
arrows. Right frame: Resonant current as a function of ac field
amplitude and energy detuning at $f=27$ GHz with $I_0 \approx 1$
pA.}\label{ener}
\end{figure}

\textit{AC-induced resonant current--} In the present work we
consider two tunnel-coupled quantum dots in the spin blockade
regime~\cite{ono02}. The quantum dots are coupled to metallic
leads, so when the spin blockade is lifted electrical current can
flow through the DQD. The physical system is described using the
same quantum transport model as in Ref.~\onlinecite{giavaras19b}.
We assume that an ac electric field modulates periodically the
energy detuning of the DQD, namely,
\begin{equation}\label{detun}
\delta(t) = - \varepsilon + A\cos(2\pi f t).
\end{equation}
Without the driving field ($A=0$) the time independent parameter
$\varepsilon$ controls the character of the singlet states as well
as the singlet-triplet spacing. When $A\ne0$ and for the
appropriate frequencies electrically driven spin-resonances can be
induced and a resonant current flows through the DQD. For large
driving amplitudes LZSM dynamics becomes relevant resulting in
interference fringes~\cite{shevchenko1, shevchenko2}. The DQD
Hamiltonian in the singlet-triplet basis: $|S_{11}\rangle$,
$|T_{+}\rangle$, $|S_{02}\rangle$, $|T_{-}\rangle$, and
$|T_{0}\rangle$ is given in matrix form~\cite{giavaras19b}
\begin{equation}\label{hamilton}
H =\left(%
\begin{array}{ccccc}
  0 & 0 & -\sqrt{2}t_{\mathrm{c}} & 0 & \Delta^{-} \\
  0 & -\Delta^{+} & -t_{\mathrm{so}} & 0 & 0 \\
  -\sqrt{2}t_{\mathrm{c}} & -t_{\mathrm{so}} & \delta & -t_{\mathrm{so}} & 0 \\
  0 & 0 &  -t_{\mathrm{so}} & \Delta^{+} &0\\
  \Delta^{-} & 0 & 0 & 0 & 0 \\
\end{array}%
\right),
\end{equation}
with the time dependent detuning, $\delta=\delta(t)$, and the
Zeeman splitting terms $\Delta^{\pm}=(\Delta_2 \pm \Delta_1)/2$,
$\Delta_{i} = g_i\mu_{\rm B} B$.

\begin{figure*}
\includegraphics[width=5.cm, angle=270]{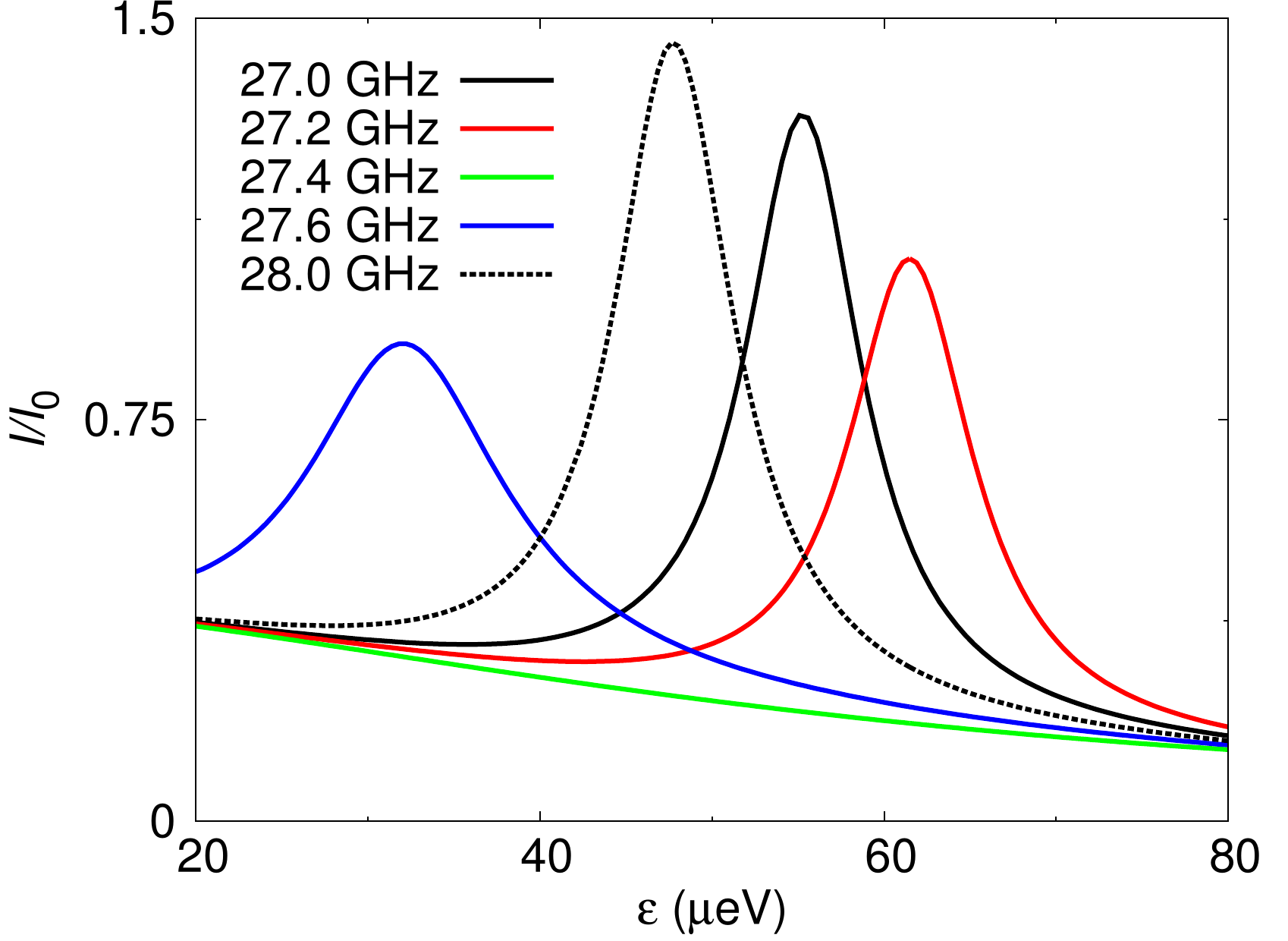}
\includegraphics[width=5.cm, angle=270]{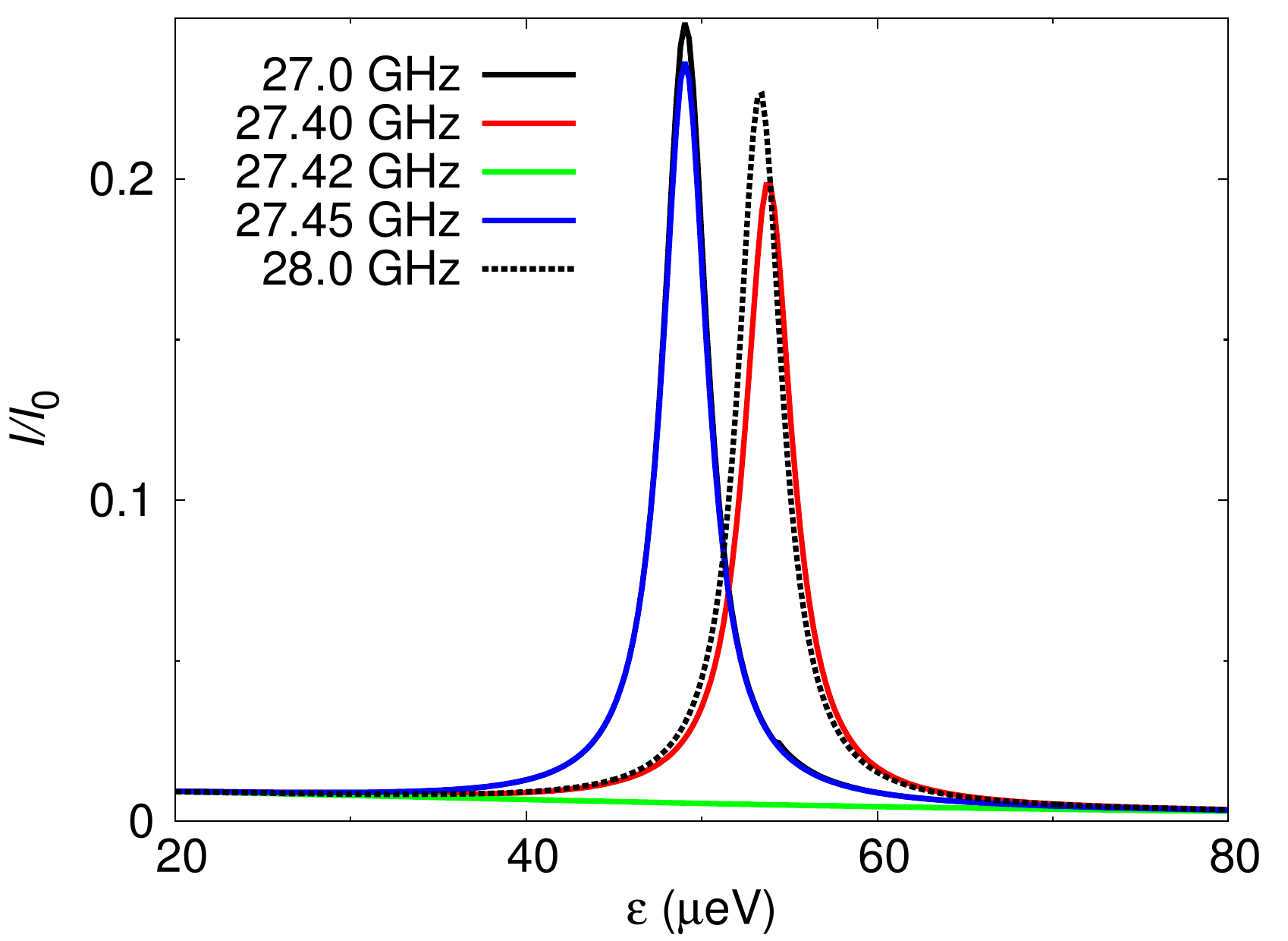}\\
\includegraphics[width=5.cm, angle=270]{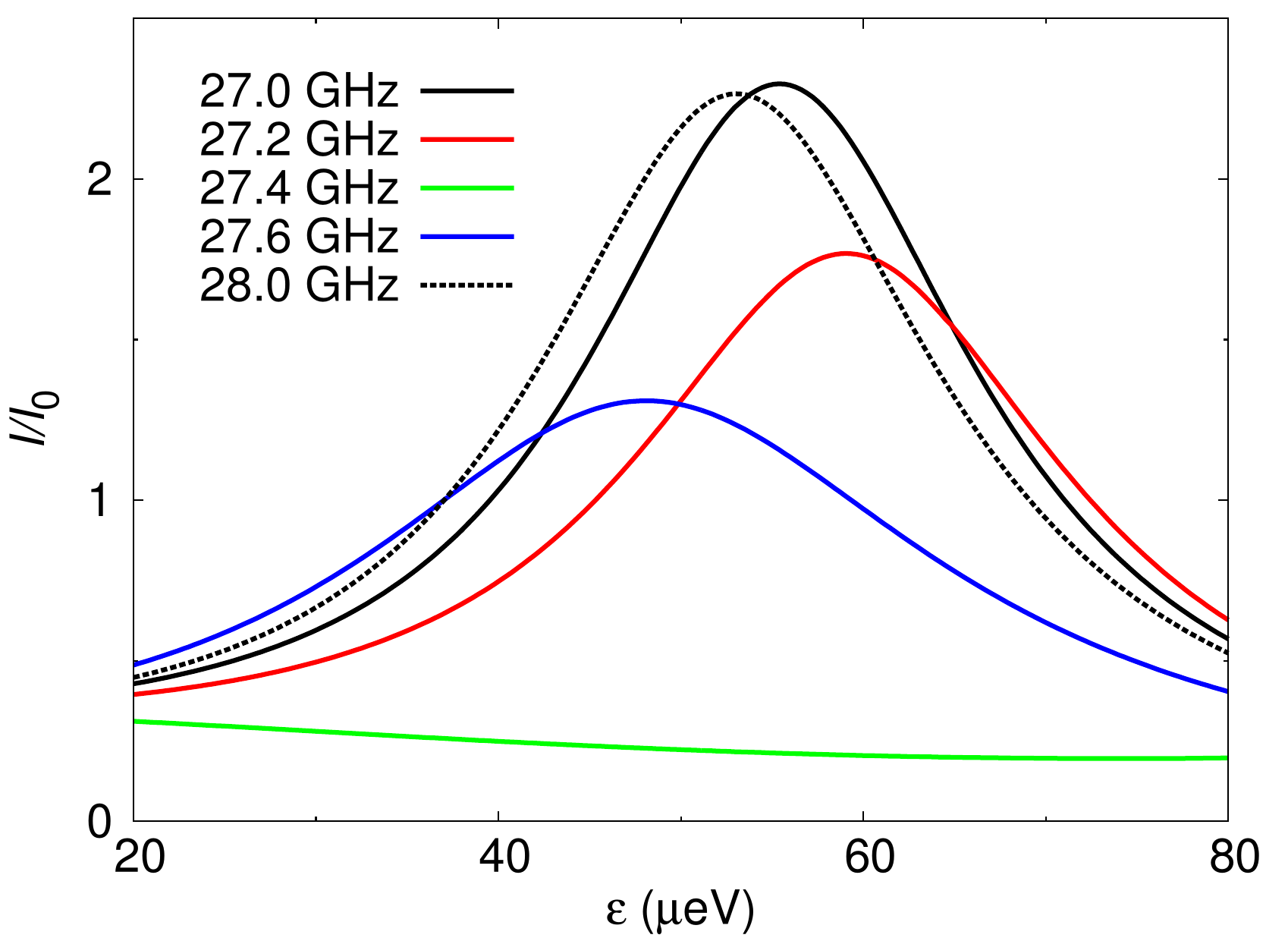}
\includegraphics[width=5.cm, angle=270]{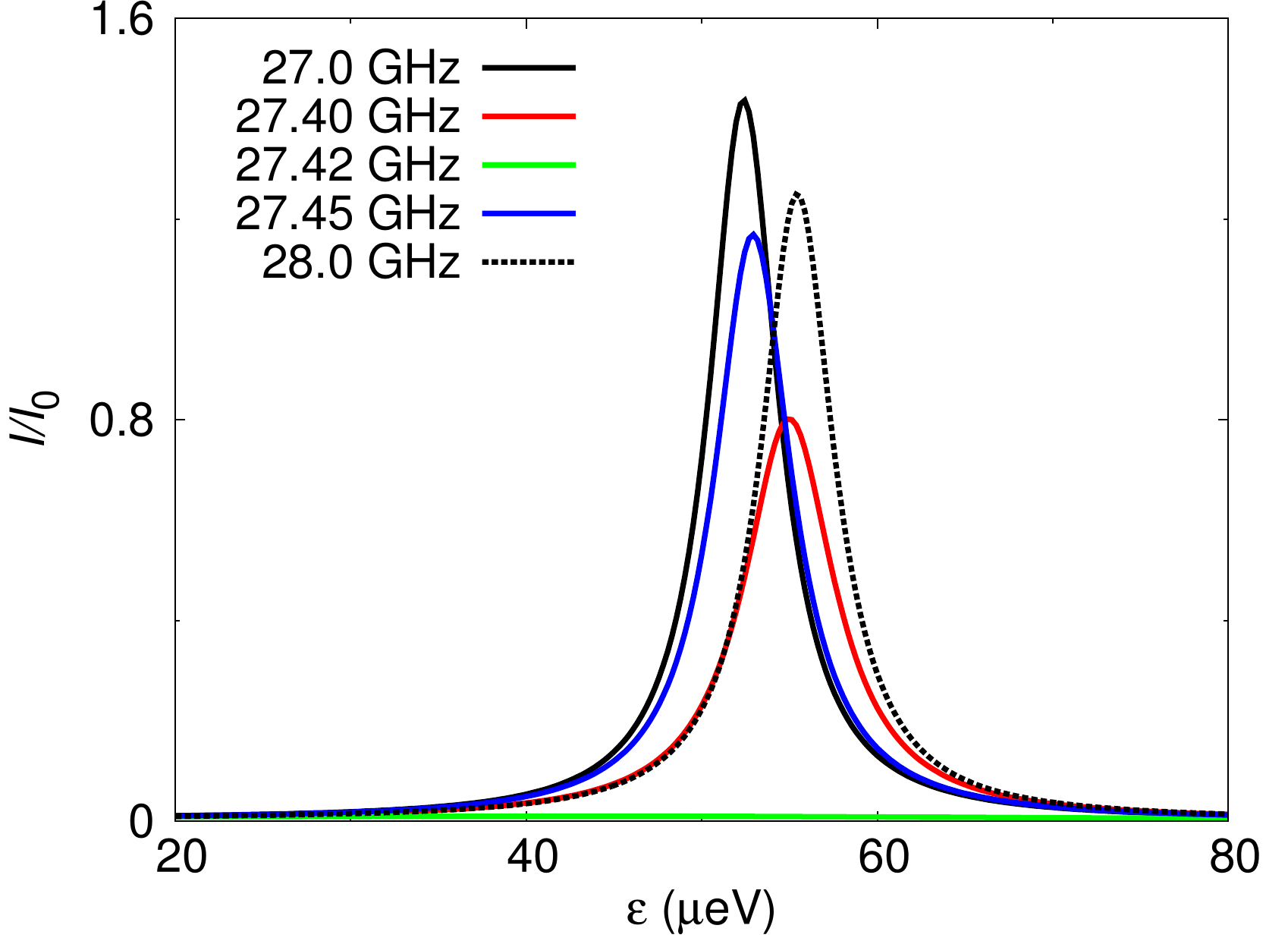}
\caption{Resonant current as a function of energy detuning at
different ac field frequencies with $t_{\rm so}= 3.5$ $\mu$eV
(left), and $t_{\rm so} = 0.5$ $\mu$eV (right). Upper frames:
$A=20$ $\mu$eV. Lower frames: $A=100$ $\mu$eV.}\label{cur}
\end{figure*}

In the spin blockade regime the two lowest singlet states of the
DQD form an anticrossing point for $\varepsilon=0$ due to coherent
interdot tunneling ($t_c\ne0$). In this tunneling process the spin
is conserved. Tuning $\varepsilon$ changes the character of the
singlet states from $(1, 1)$ to $(0, 2)$ where the notation $(n,
m)$ indicates $n$ ($m$) electrons in the left (right) quantum dot.
In addition, the two lowest singlet states and the $S_z=\pm 1$
triplet states form anticrossing points due to a nonzero
spin-orbit tunnel coupling ($t_{\rm so}\ne0$). This tunneling
involves a coherent spin flip forming hybridized singlet-triplet
states. For such a DQD system an earlier work~\cite{giavaras19b}
has explored the ac-induced current as a function of the driving
amplitude and demonstrated the significant role of $t_{\rm so}$ in
the interference profile. Most importantly ac-induced blocked
(dark) states suppress the resonant current at very characteristic
values of the ratio $A/hf$~\cite{giavaras19b}. Despite the fact
that Ref.~\onlinecite{giavaras19b} focuses on a narrow energy
detuning range for $\varepsilon>0$, the results can be
straightforwardly extended to larger positive detunings as well as
to $\varepsilon \le 0$. In this case interference fringes can be
observed which resemble those of a two-level quantum system.

\begin{figure*}
\includegraphics[width=3.8cm, angle=270]{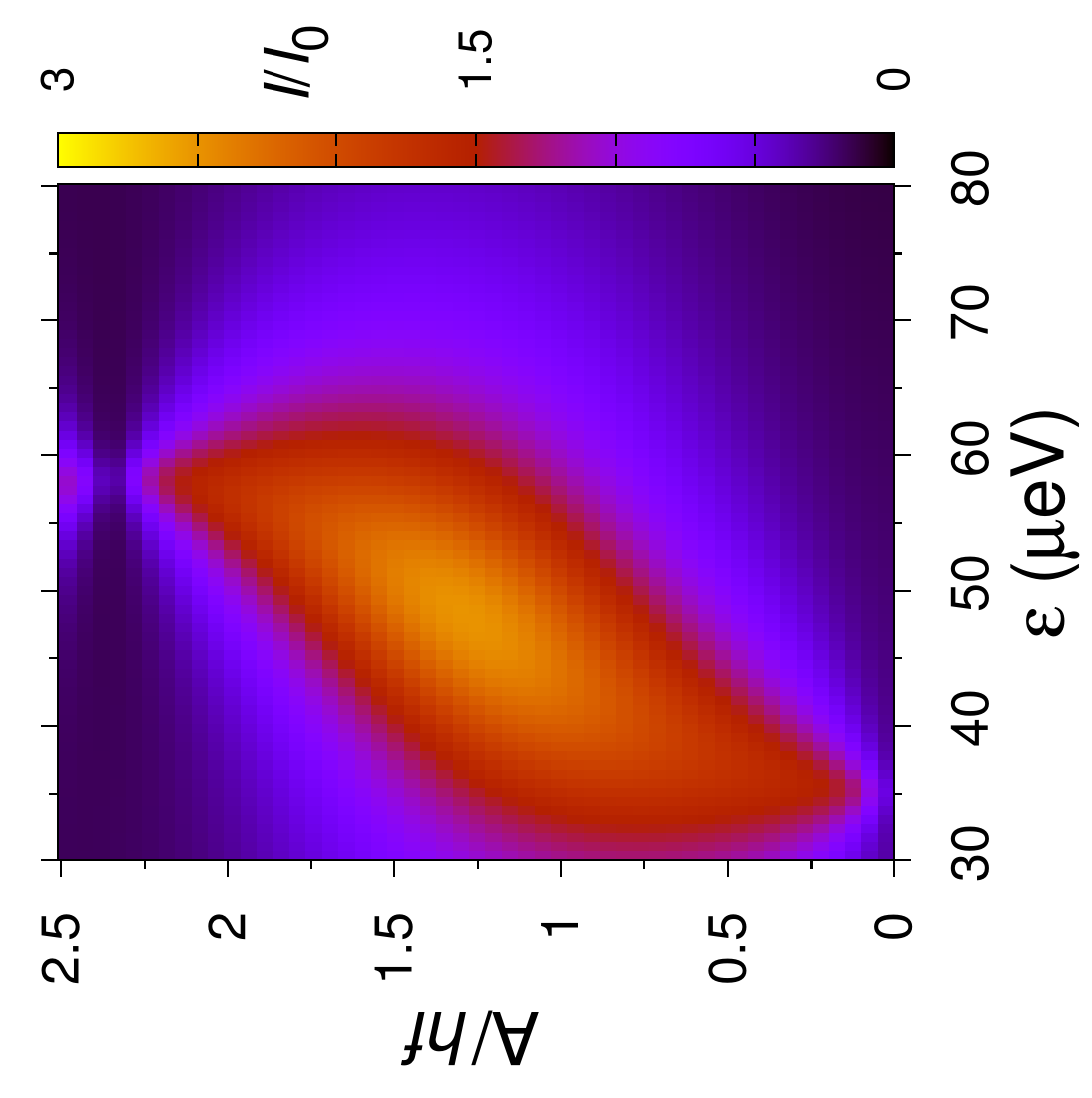}
\includegraphics[width=3.8cm, angle=270]{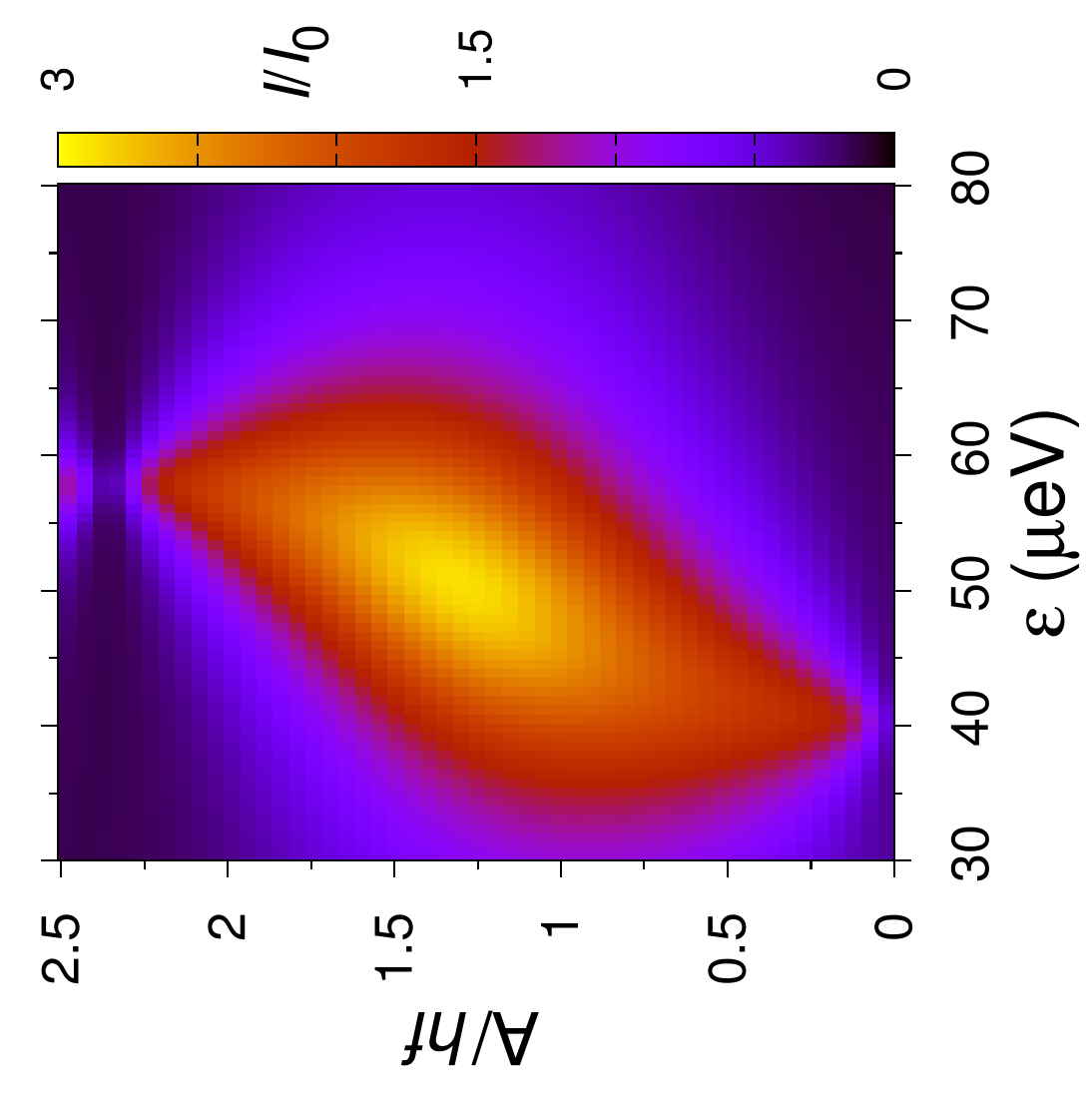}
\includegraphics[width=3.8cm, angle=270]{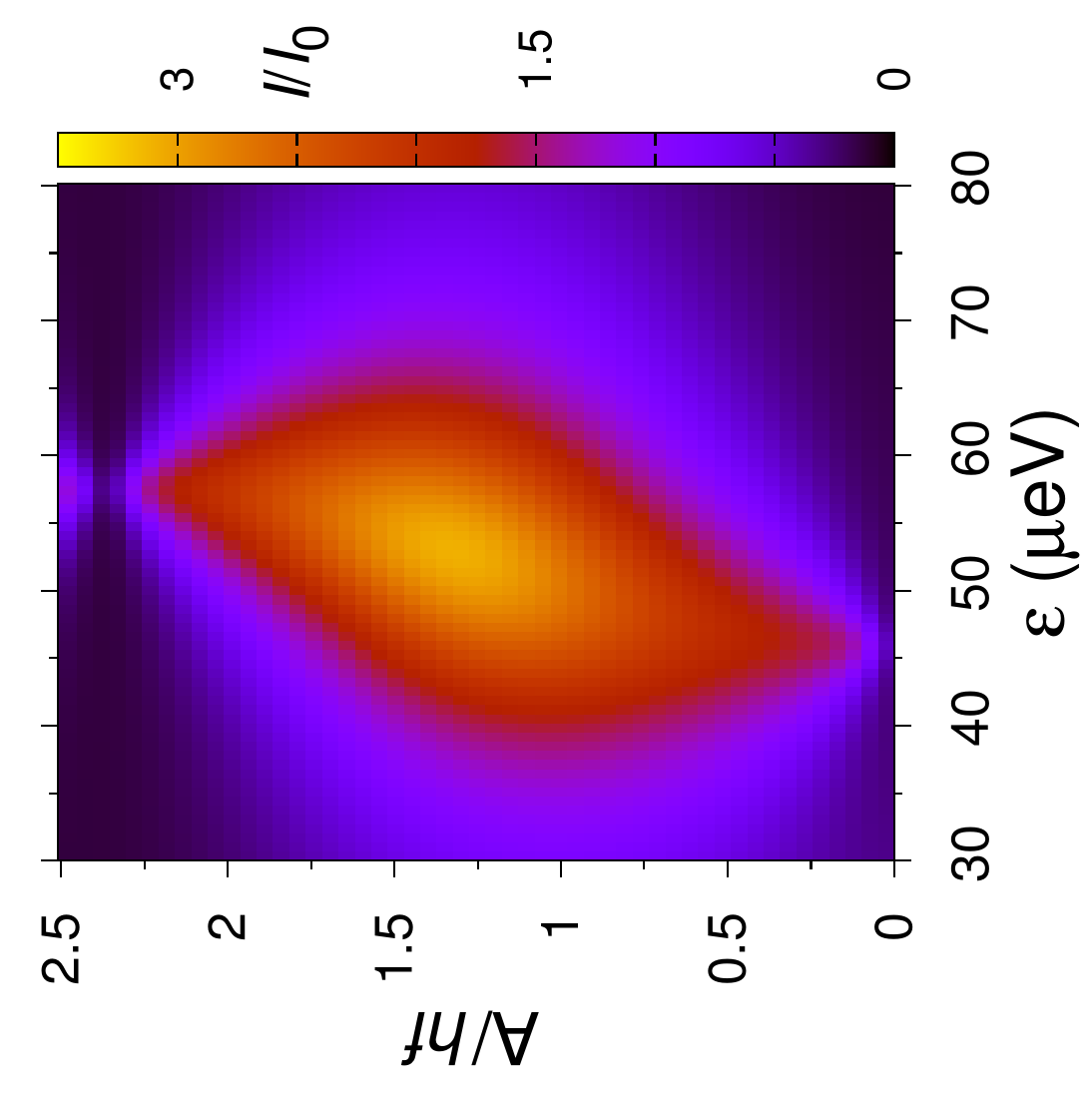}
\includegraphics[width=3.8cm, angle=270]{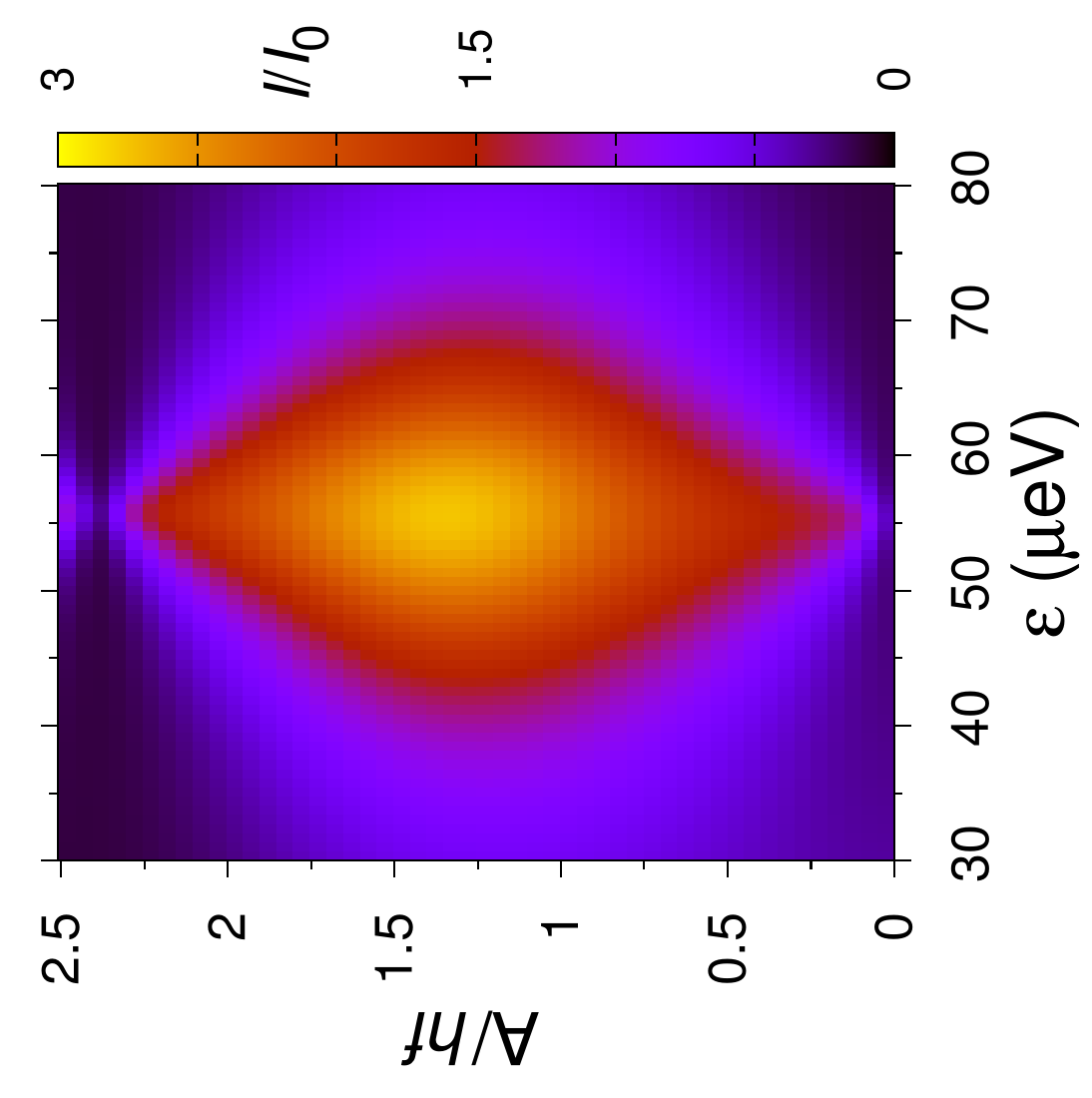}\\
\includegraphics[width=3.8cm, angle=270]{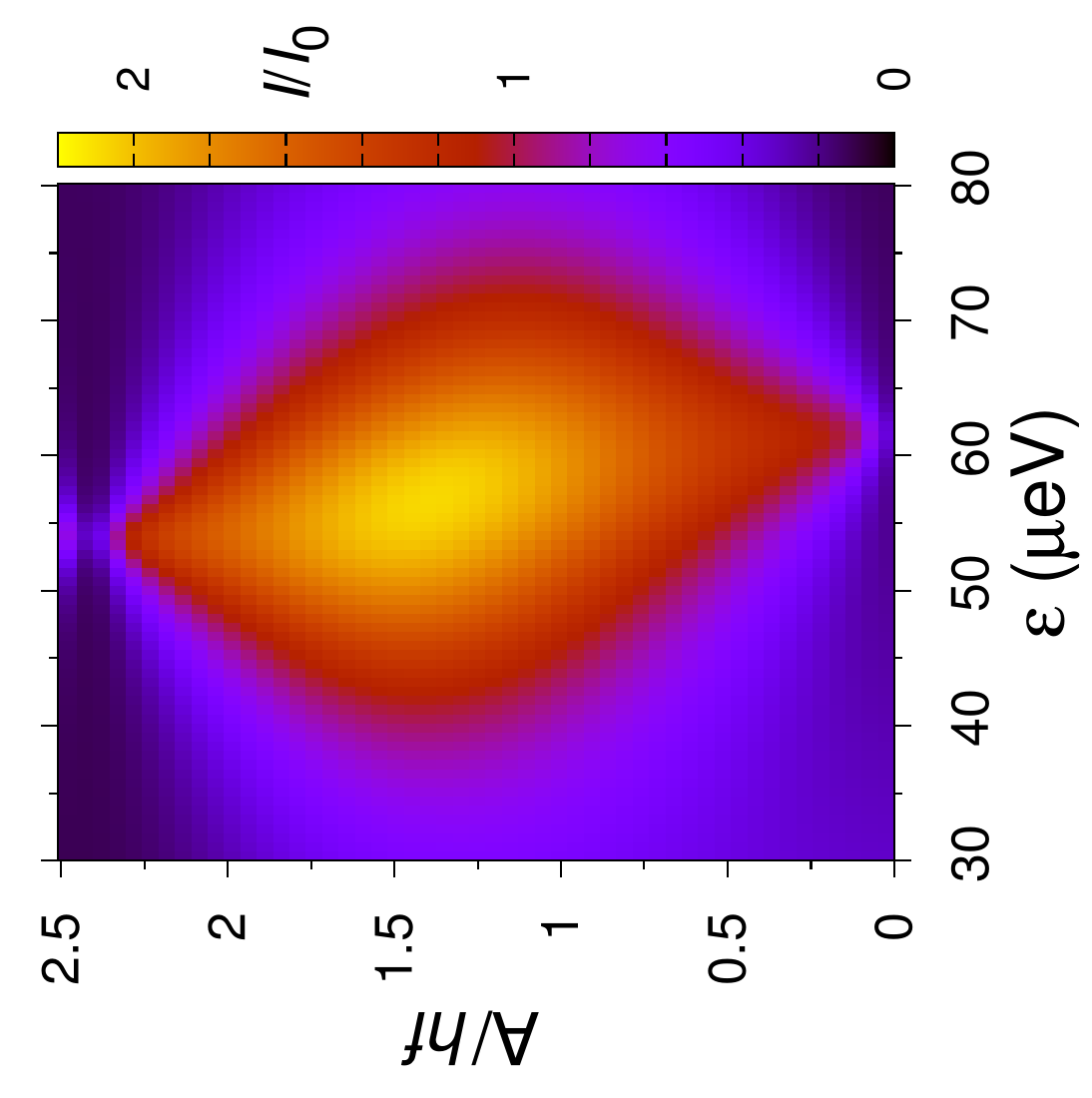}
\includegraphics[width=3.8cm, angle=270]{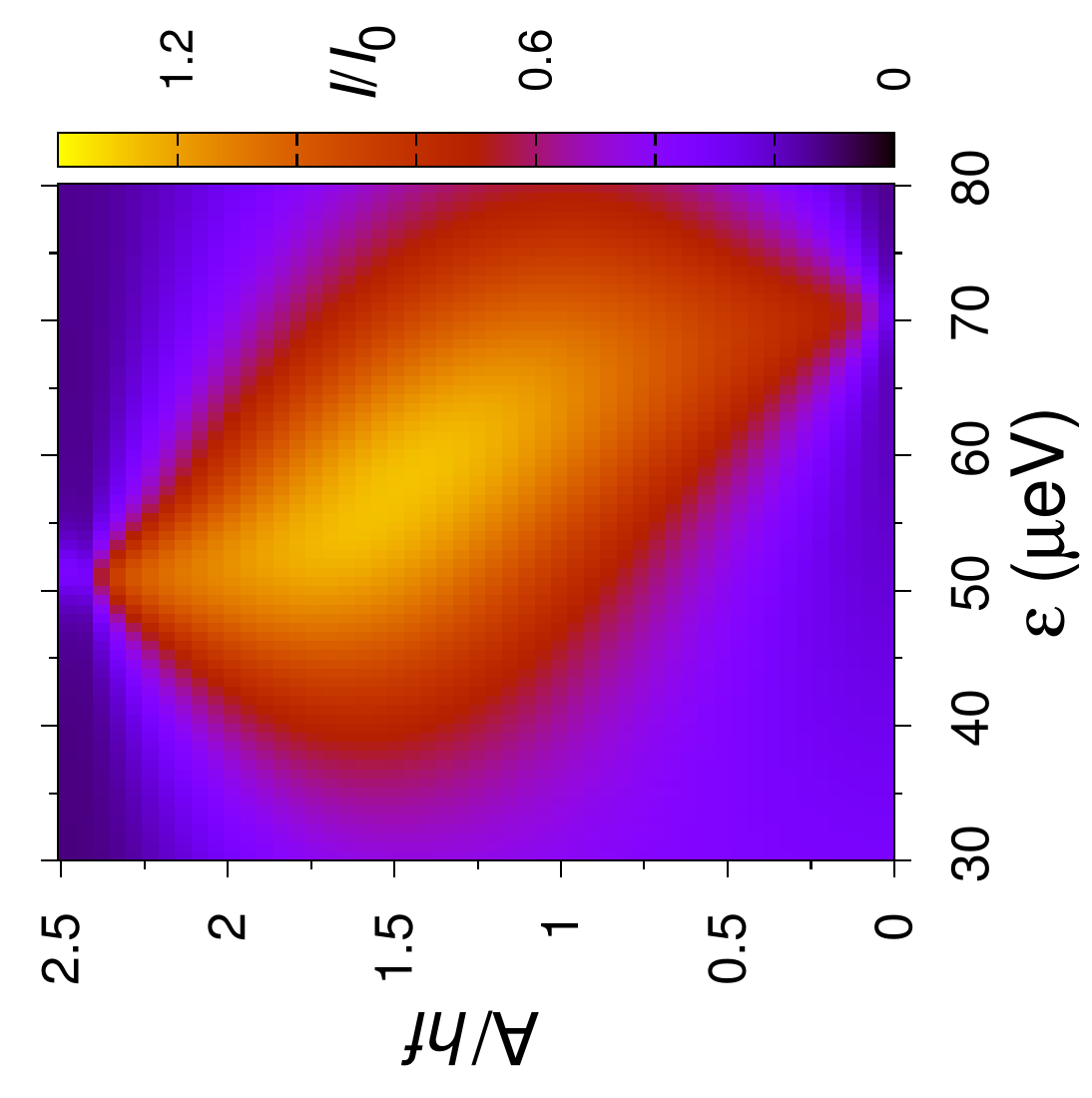}
\includegraphics[width=3.8cm, angle=270]{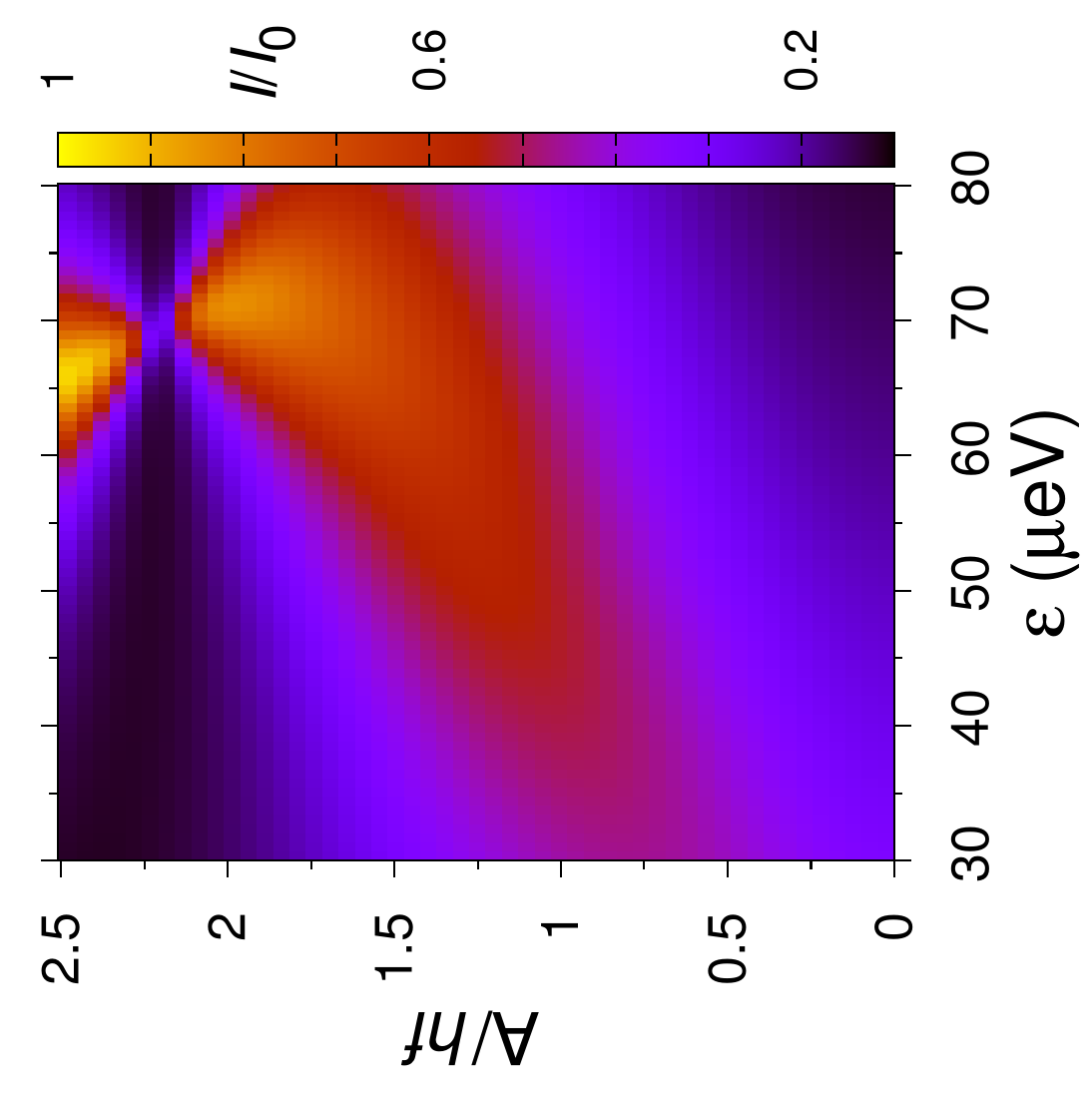}
\includegraphics[width=3.8cm, angle=270]{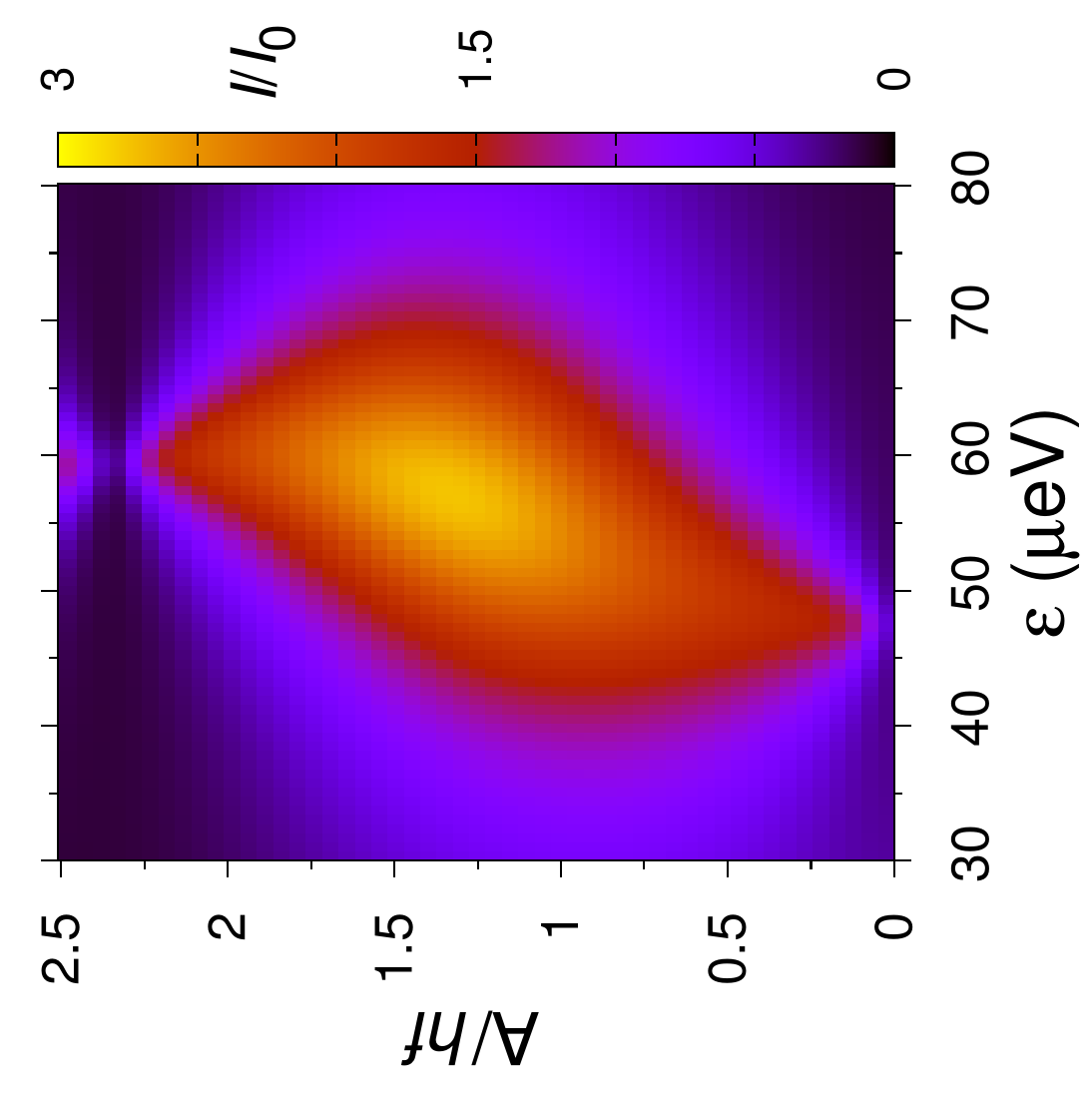}
\caption{Resonant current as a function of ac field amplitude and
energy detuning for $t_{\rm so}= 3.5$ $\mu$eV. The ac field
frequency in GHz is: $f=24$, 25, 26, 27 (upper frames) and 27.2,
27.3, 27.5, 28 (lower frames).}\label{plot1}
\end{figure*}

The eigenenergies ($A=0$) of the DQD as a function of the detuning
are plotted in Fig.~\ref{ener}. At $\varepsilon \approx 50$ meV
the energy level $E_5$ corresponds to a triplet-like state,
whereas the levels $E_1$ and $E_2$ correspond to hybridised
singlet-triplet states due to $t_{\rm so} \ne 0$. In the present
work typical frequencies are of the order of $hf \sim E_{5}-E_{1}
\approx E_{5} - E_{2}$ and the resonant current is computed for
small values of $\varepsilon$ near zero. Larger positive detunings
as well as negative values can equally be considered. The resonant
current is obtained from the steady-state density matrix of the
DQD. This density matrix obeys a Floquet-Markov master equation of
motion~\cite{flqmas1} in the sequential tunneling regime and with
the same periodicity as that of the ac field. This approach is
more appropriate in order to capture strong driving effects,
however, approximate treatments can provide valuable
insight~\cite{giavaras19a, giavaras19b, giavaras20} even in the
`multi-photon' regime.

Some characteristic examples of the resonant current as a function
of the energy detuning are shown in Fig.~\ref{cur}. Similar to an
earlier study~\cite{giavaras19b} we choose the ac field frequency
to be of the order of $hf \sim E_{5}-E_{1} \approx E_5 - E_2$;
this means that we tune the frequency so that to capture the
anticrossing point between the levels $E_1$ and $E_2$ shown in
Fig.~1. To clarify the role of the driving amplitude we compute
the current for two different cases. When $A$ is chosen to be
relatively small, i.e., $A\approx 20$ $\mu$eV, the induced current
peaks are indicative of singlet-triplet transitions which lift the
spin blockade. However, increasing the driving amplitude gradually
introduces deviations from this simple picture. This can be seen
for $A \approx 100$ $\mu$eV where the current peaks start to
overlap. According to Fig.~\ref{cur}, small changes in $f$ near
the anticrossing lead to a noticeable current suppression; in
agreement with an earlier work~\cite{giavaras19b}. Specifically,
the current peak is nearly vanished and the observed current is
approximately equal to the current flowing through the DQD without
the ac field ($A=0$). This current is computed with the same model
as in Ref.~\onlinecite{giavaras13} and is always nonzero due to
the inclusion of the single electron states in the equation of
motion of the DQD density matrix. These lead to non-zero matrix
elements between states differing by one electron, thus, lifting
the ideal spin blockade condition~\cite{giavaras13}.

A simple investigation shows that the current suppression in
Fig.~\ref{cur} marks the singlet-triplet anticrossing.
Consequently, a smaller spin-orbit tunnel coupling (leading to a
smaller anticrossing gap) requires a finer $f$-tuning for the
current suppression to be observable. This remark is consistent
with the results in Fig.~\ref{cur} that correspond to $B=0.135$ T.
Decreasing the magnetic field gradually changes the character of
the two-spin states, therefore deviations from the physical
picture presented in Fig.~\ref{cur} are expected especially at
very low magnetic fields. In contrast, similar trends are observed
at higher fields.

The results in Fig.~\ref{cur} can be extended to larger values of
$A$ allowing us to go beyond the simple spectroscopic regime. An
important aspect that complements the findings of an earlier
work~\cite{giavaras19b} is that the resonant current exhibits an
interesting interference profile even for a (relatively) small
amplitude to frequency ratio, i.e., $A / hf < 3$. To quantify this
argument we present some results in Fig.~\ref{plot1} for different
values of $f$. These reveal a high-current region that has similar
characteristics to the region that has been predicted in an
earlier study~\cite{giavaras19b}. Increasing $f$ leads to small
changes in the resonant current, however, as $f$ approaches the
singlet-triplet anticrossing [ $h f\approx (g_1+g_2)\mu_{\rm B}
B$] the overall current profile becomes very sensitive to the
exact value of $f$. Simultaneously a current suppression is
observed, especially for small ratios $A/hf$, and a drastic change
occurs in the resonant region.

Although in Fig.~\ref{plot1} we focus on $\varepsilon>0$ and near
zero only, the above current characteristics can also be found for
negative values of $\varepsilon$. In addition, increasing the
ratio $A / hf$ allows us to explore in more detail the resulting
interference fringes (similar to Fig.~1) due to the LZMS dynamics.
Finally, we remark that using a time dependent gate voltage to
periodically control the energy detuning [Eq.~(\ref{detun})] is
very likely to unintentionally induce a weak time dependence in
the interdot tunnel coupling, $t_{c}$ and/or $t_{\rm so}$. This is
probable to occur for large driving amplitudes, so for this reason
a study of a realistic DQD should account for a weak time
dependent tunnel coupling term. Theoretical
studies~\cite{giavaras19a, giavaras20} have demonstrated that such
a term gives rise to a resonant current that can dominate over the
current in Fig.~\ref{plot1}.

\textit{Summary--} In summary, we computed the resonant current in
tunnel coupled quantum dots for different ratios $A/hf$ and energy
detunings. We found that beyond the spectroscopic regime which
reveals the anticrossing gap the resonant current exhibits a rich
interference pattern. We showed that by considering a narrow
detuning range and investigating the current at different ac field
frequencies can give insight into a singlet-triplet state mixing.
Our results can be validated with electrical transport
measurements.

\end{document}